\documentclass[a4paper,10pt,twoside]{cpc-hepnp}

\usepackage{multicol}
\usepackage{graphicx}
\usepackage{booktabs}
\usepackage{amssymb,bm,mathrsfs,bbm,amscd}
\usepackage[tbtags]{amsmath}
\usepackage{lastpage}

\begin{document}

\fancyhead[c]{Submitted to `Chinese Physics C'}

\title{Generation and measurement of sub-picosecond electron bunch in photocathode rf gun\thanks{Supported by National Natural Science Foundation of China (11205152) and Science Foundation of Ministry of Education of China (``985 project'': 173123200402002) }}

\author{%
      LI Wei-Wei
\quad HE Zhi-Gang$^{1)}$\email{hezhg@ustc.edu.cn}%
%\quad WANG Xiao-Jun(ÍõС¾ü)$^{1,2;2)}$\email{hepnp@mail.ihep.ac.cn}%
\quad JIA Qi-Ka} \maketitle

\address{%
National Synchrotron Radiation Laboratory, University of Science and
Technology of China, Hefei, 230029, Anhui, China\\
%$^2$ {\bf Example}: Institute of High Energy Physics, Chinese Academy of Sciences, Beijing 100049, China\\
}

\begin{abstract}
We consider a scheme to generate sub-picosecond electron bunch in the photocathode rf gun by improving the acceleration gradient in the gun, suitably tuning the bunch charge, the laser spot size and the acceleration phase, and reducing the growth of transverse emittance by laser shaping. A nondestructive technique is also reported to measure the electron bunch length, by measuring the high-frequency spectrum of wakefield radiation which is caused by the passage of a relativistic electron bunch through a channel surrounded by a dielectric.
\end{abstract}

\begin{keyword}
photocathode rf gun, sub-picosecond electron bunch, bunch length measurement, wakefield radiation
\end{keyword}

\begin{pacs}
29.25.Bx, 29.27.Ac, 29.27.Fh
\end{pacs}

\begin{multicols}{2}

\section{Introduction}

Sub-picosecond electron bunch is generally required in lots of applications, such as the generation of  coherent synchrotron radiation\cite{lab1}, coherent Smith-Purcell radiation\cite{lab2} and coherent cherenkov radiation\cite{lab3} whose spectra are in the THz gap; the production of sub-picosecond X-ray pulse by Compton backscattering\cite{lab4}. For the time-resolved MeV ultra-fast electron diffraction\cite{lab5}, the length of electron bunch is also needed to be sub-picosecond and as shorter as possible to achieve higher time resolution.

Magnetic compression with chicane and velocity bunching in a linear accelerator are customary and effective ways used to get a short bunch. However both of the two techniques take large space, while the table-top experimental facility is needed in some applications. In this paper, we analyse the impact factors of the electron bunch length, and consider a scheme to directly generate the sub-picosecond electron bunch in the photocathode rf gun by improving the acceleration gradient in the gun, suitably tuning the bunch charge, the laser spot size and the acceleration phase, and reducing the growth of transverse emittance by laser shaping.

In respect of the bunch length measurement, there are many methods, such as rf deflecting cavity\cite{lab6}, rf zero phasing\cite{lab7}, electro-optical sampling\cite{lab8}, coherent radiation\cite{lab9,lab10}, and so on. Each diagnostic has its advantages and disadvantages. Ideally, one would want a diagnostic that disturbs the bunch as little as possible, can be single shot measurement, and uses instrumentation that is inexpensive, easy to adjust, and routine to calibrate. A nondestructive bunch length diagnostic technique was reported in Reference\cite{lab11}, which can measure the rms bunch length by observing the frequency spectrum of wakefield radiation as the bunch passes through a vacuum channel in a hollow dielectric element. The design of a bunch length measurement system based on this technique is reported in this paper.

\section{Generation of sub-picosecond electron bunch}

The length of electron bunch is affected by such factors as space charge effect, beam energy and energy spread, and the coupling effect between the transverse and longitudinal emittances.

The bunch lengthening due to the space charge effect can be estimated in a drift space by\cite{lab12}:
\begin{equation}
\Delta\sigma_z=2qcL^2/I_aR\sigma_z\gamma^4
\end{equation}
where q is the charge of bunch, c is the speed of light, L is the drift distance, ~$I_a=1.7~kA$~, R is the bunch radius, ~$\sigma_z$~ is the bunch length and ~$\gamma$~ is the beam
energy. In the photocathode rf gun, the energy of electron beam is low, so the space charge effect plays the dominant role. In order to decrease the bunch lengthening caused by the space charge effect, the acceleration gradient should be as high as possible and the bunch charge should be chosen appropriately. Furthermore, the bunch length can be compressed in the gun by tuning the acceleration phase\cite{lab13}. For our laser pulse (the measured rms length is about 2.0 ps), we use the code ASTRA\cite{lab14} to simulate the bunch length evolution as a function of acceleration phase at different charges and acceleration gradients, and the results are shown in Fig. 1.
\begin{center}
    \includegraphics[width=8.0cm]{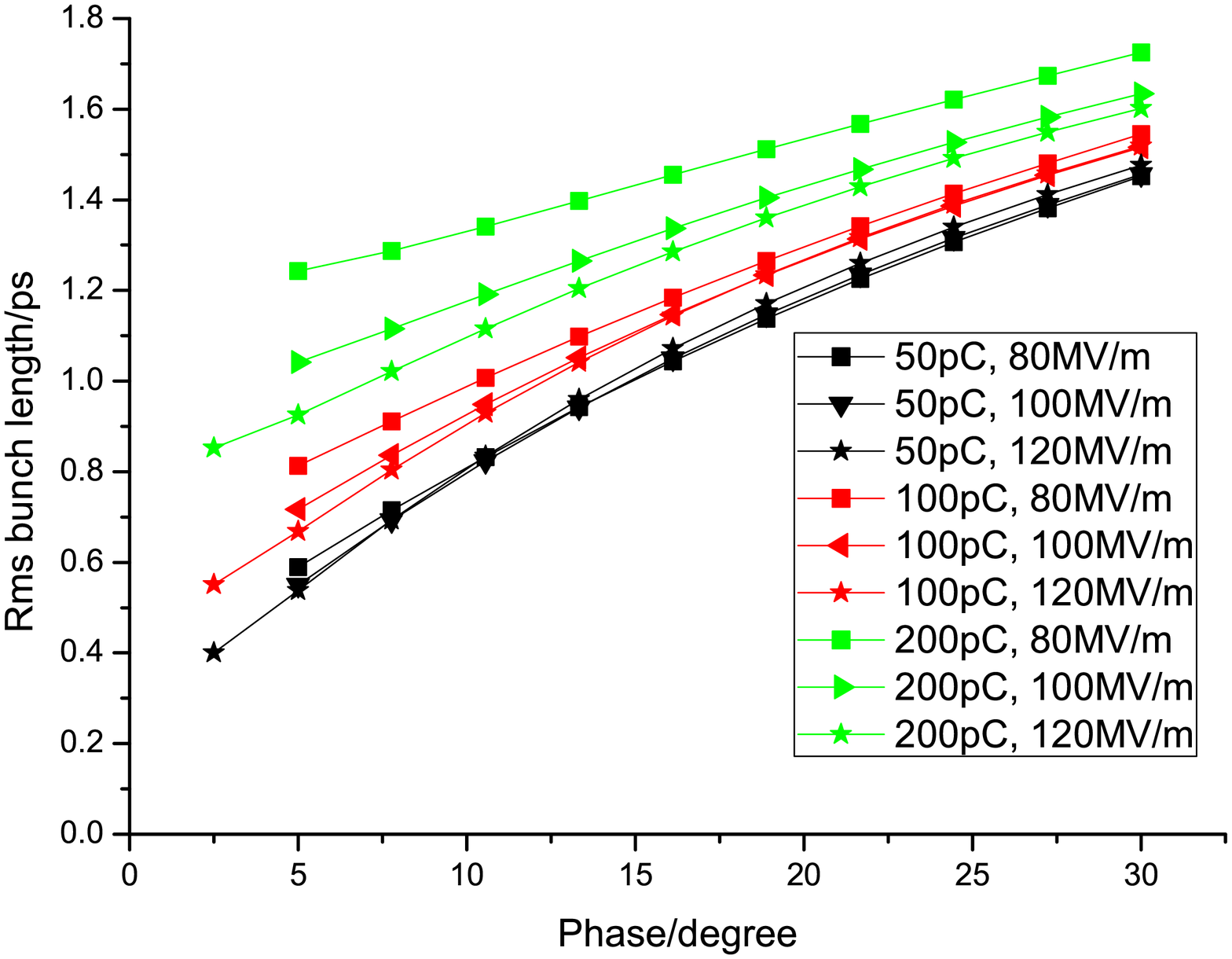}
    \figcaption{\label{fig1} The rms bunch length vs. acceleration phase at different charges and acceleration gradients.}
    \end{center}
The acceleration gradient of our photocathode rf gun, which is a second generation gun machined by the Department of Engineering Physics of Tsinghua University, is achieved at about 80 MV/m. The acceleration gradient of the third generation gun is achieved at 120 MV/m at present\cite{lab15}, whose cathode seal technique is improved  by replacing the HELICOFLEX seal with a MATSUMOTO gasket to eliminate the cathode gap as much as possible\cite{lab16}.

The bunch length can be compressed further in the drift space by a suitable energy spread of the beam, if the length of drive laser pulse is appropriate\cite{lab17}. For our relatively short laser pulse, the bunch length changing caused by energy spread is small because of the relatively short drift space and the small energy spread at the exit of the gun.

The coupling between the transverse and longitudinal emittances is another factor in lengthening the electron bunch. The laser shaping technique is an effective way to restrain the growth of transverse emittance, which consists of spatial and temporal shaping\cite{lab18}. In this paper, we only consider the spatial shaping. Although the uniform laser spot can be achieved by using a spatial shaper, it is difficult to transport the shaped spot to the cathode. So we plan to clip the laser spot by an aperture, as shown in Fig. 2.
    \begin{center}
    \includegraphics[width=5.0cm]{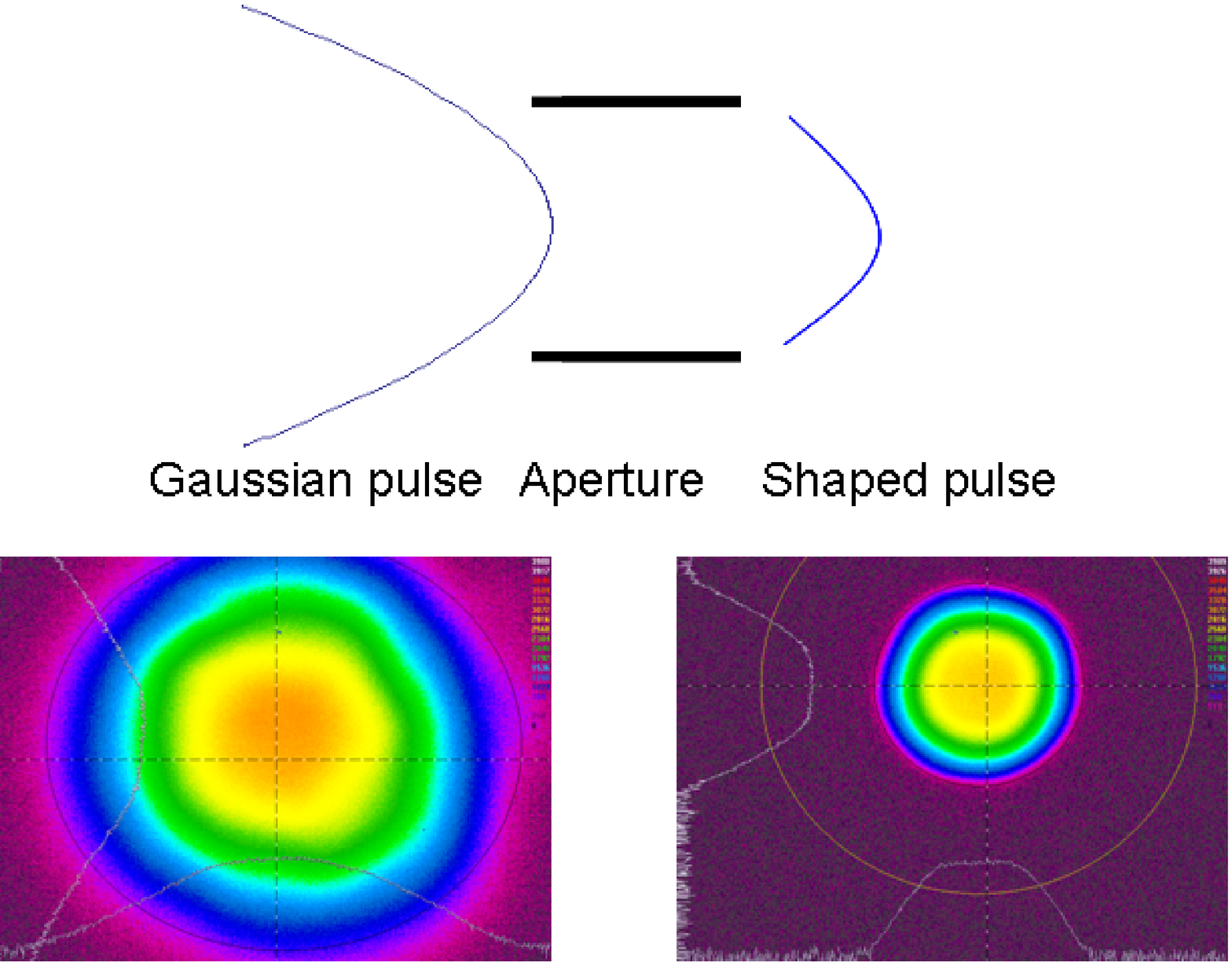}
    \figcaption{\label{fig2} Clipping shaping: sketch and shaping result (UV light)}
    \end{center}
For the 2 mm diameter gaussian spot and clipped spot (a gaussian spot with 0.8 mm rms size clipped by a 2 mm diameter aperture), the evolution of transverse emittance and rms bunch length are shown as Fig. 3, where the acceleration gradient is 120 MV/m, the bunch charge is 200 pC, the acceleration phase is 5 degrees, and the magnetic field of the solenoid is 2500 Gauss. The rms bunch length at the focal point of the solenoid (around 0.9 m) is about 0.97 ps for the clipped laser spot, while it is 1.26 ps for the gaussian laser spot.
    \begin{center}
    \includegraphics[width=8.0cm]{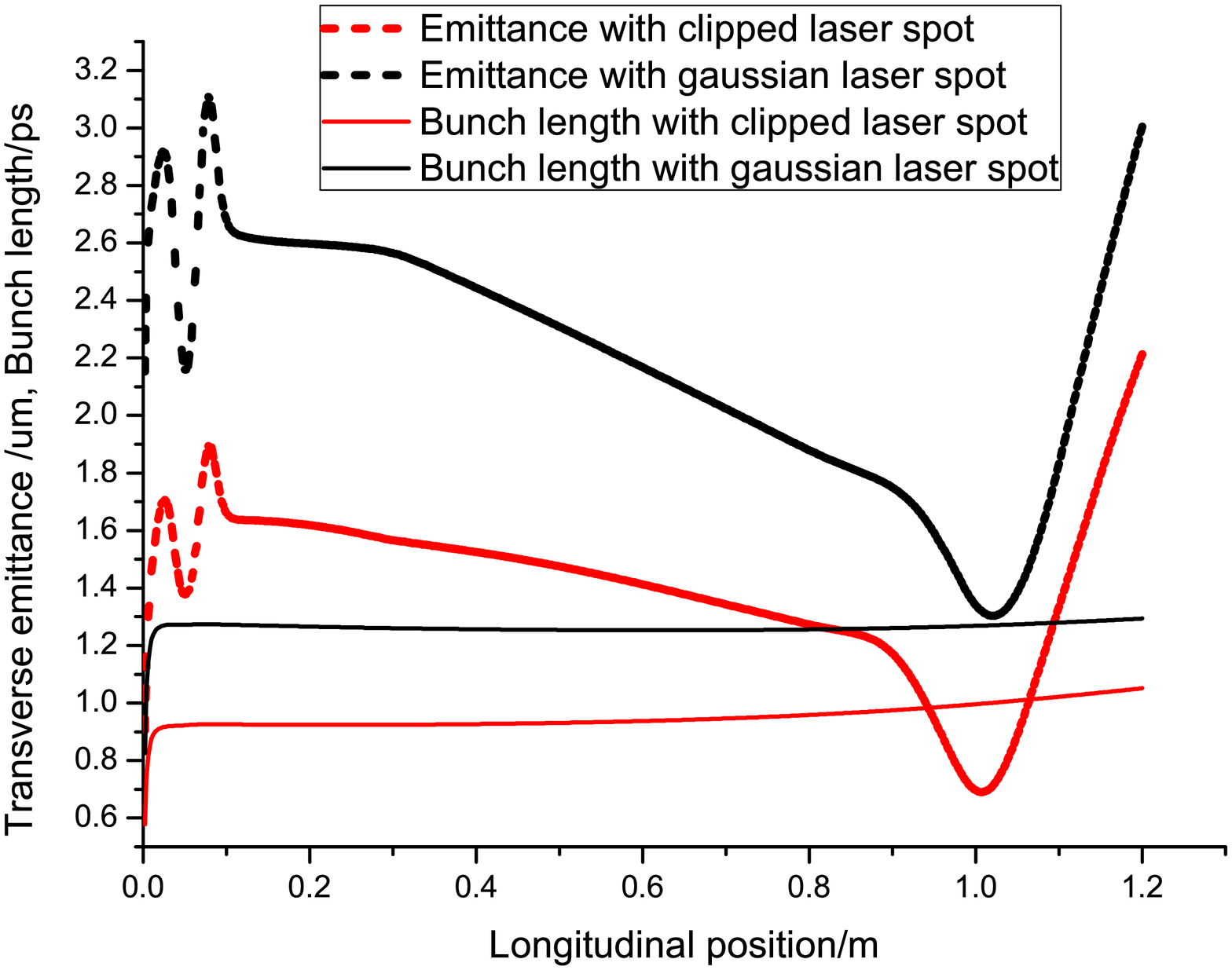}
    \figcaption{\label{fig3} Transverse emittance and rms bunch length vs longitudinal position for different laser spots}
    \end{center}

Equation (1) shows that the bunch lengthening due to the space charge effect is proportional to the diameter of laser spot. So the bunch length can be modulated by tuning the diameter of laser spot, as shown in Fig. 4, and the current distributions of bunches at the focal point of the solenoid are shown in Fig. 5. However the transverse emittance will grow seriously when the diameter of laser spot is too large. The optimal transverse emittances are ~$0.7~mm\cdot mrad$~, ~$1.2~mm\cdot mrad$~and~$1.75~mm\cdot mrad$~respectively. Nevertheless, the transverse emittance is not critical in all the applications.
 \begin{center}
    \includegraphics[width=8.0cm]{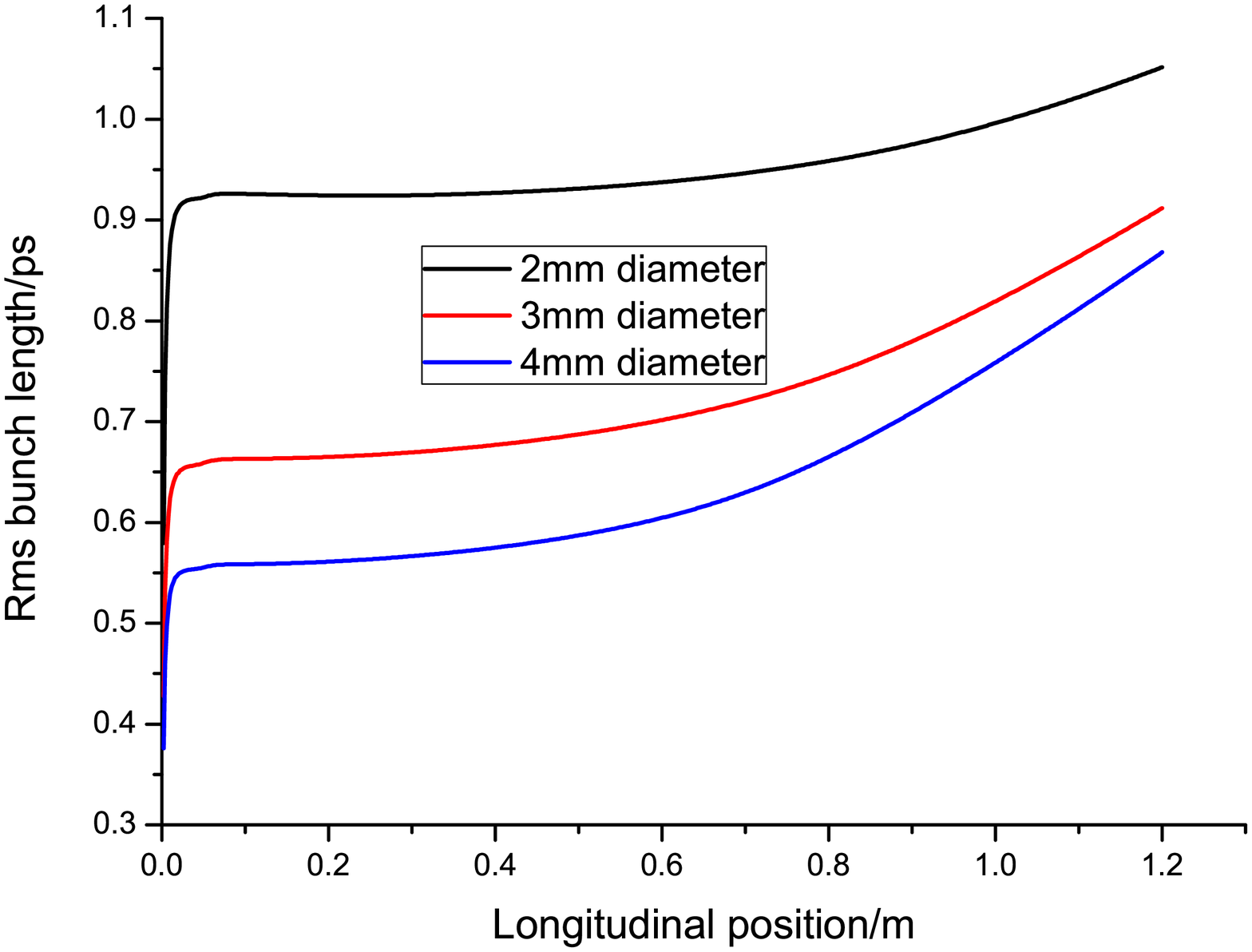}
    \figcaption{\label{fig4} The rms bunch length vs longitudinal position for clipped laser spots with different diameters}
    \end{center}

 \begin{center}
    \includegraphics[width=6.0cm]{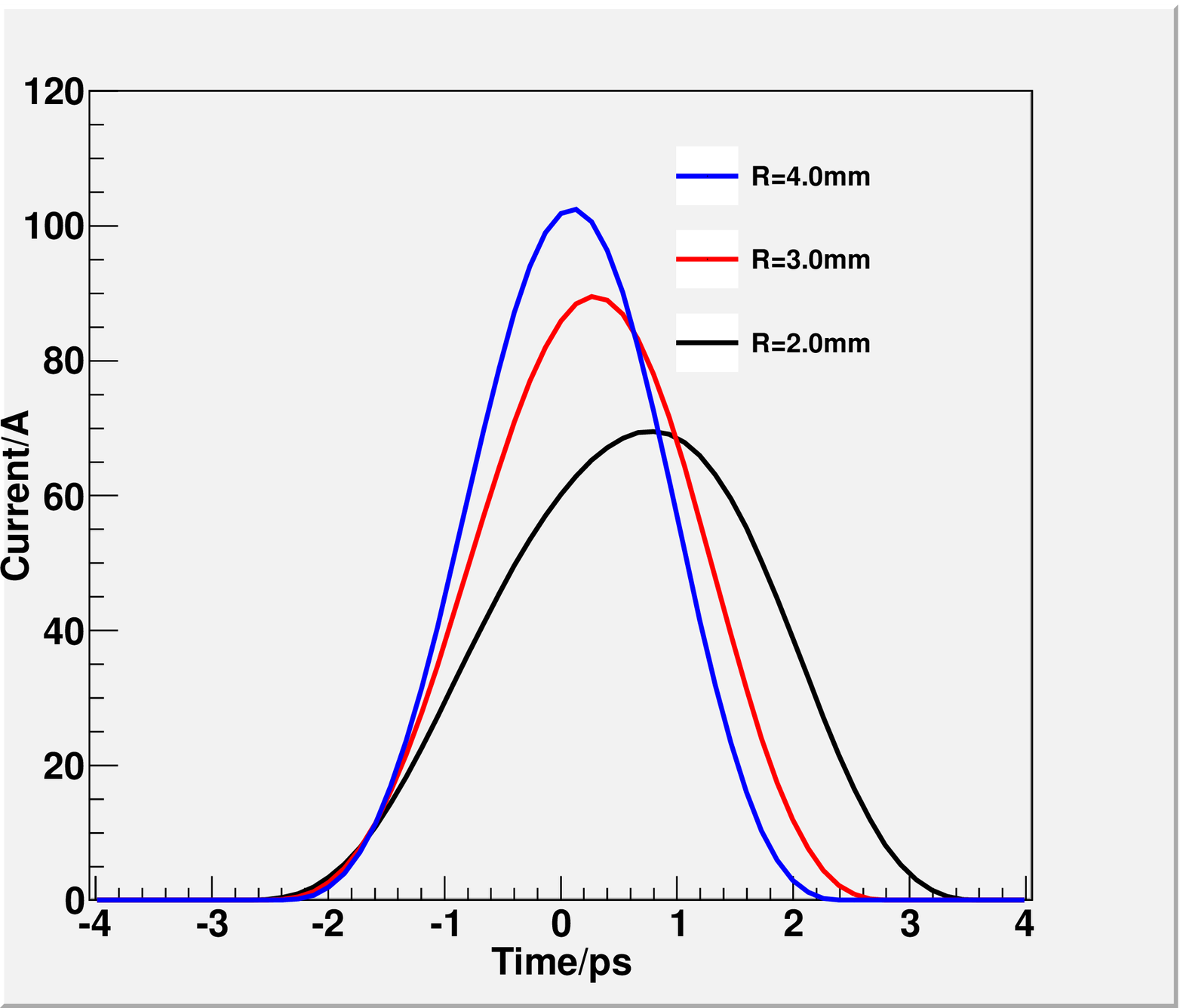}
    \figcaption{\label{fig5} Current distribution of bunches at the focal point of the solenoid }
    \end{center}
In summary, to generate sub-picosecond electron bunch directly in the photocathode rf gun, we need to improve the acceleration gradient of the gun as high as possible, restrain the growth of transverse emittance through laser shaping technique, and carefully tune the energy (bunch charge) and diameter of laser spot according to the requirements of applications.

\section{Bunch length measurement}
As a relativistic electron bunch travels along the vacuum channel in the tube, it
drives coherent Cherenkov radiation wakefields\cite{lab19} that are confined to a discrete set of modes due to the waveguide boundaries. The power (energy) at certain modes is correlative to the length of electron bunch. So the bunch length can be measured by observing the frequency spectrum of the wakefield radiation. Fig. 6 is a sectional drawing of hollow cylindrical dielectric tube coated on the outer surface with metal to form a dielectric-lined waveguide. We now present a brief summary of analysis for the
fields set up within this structure.
   \begin{center}
    \includegraphics[width=7.0cm]{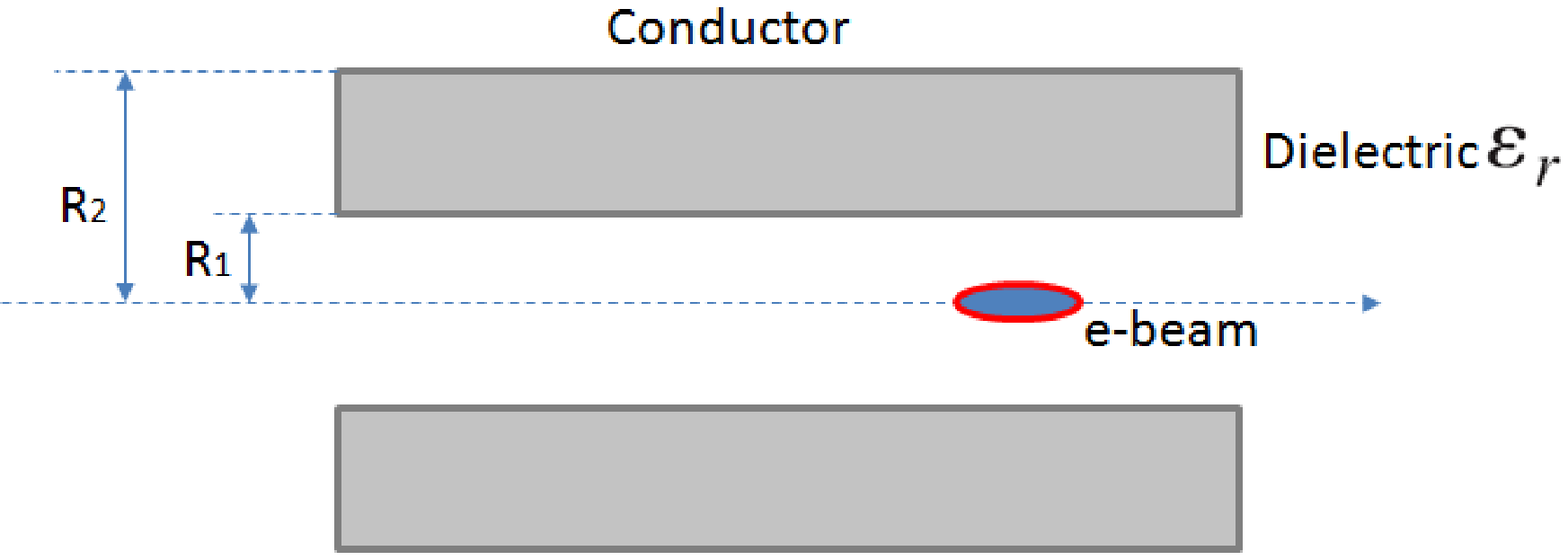}
    \figcaption{\label{fig6} The sectional drawing of a beam-driven cylindrical dielectric-lined waveguide }
    \end{center}

In the analysis, Gauss system of units is used. Fourier expansion of the longitudinal fields in a circular cylindrical waveguide takes the form:
\begin{equation}
\begin{array}{c}
 \left( {\begin{array}{*{20}{c}}
   {{E_z}(r,t)}  \\
   {{H_z}(r,t)}  \\
\end{array}} \right) = \frac{1}{{{{(2\pi )}^3}}}\int_{ - \infty }^\infty  {d\omega d} k \times  \\
 \sum\limits_{l =  - \infty }^\infty  {\exp \left[ { - i(\omega t - kz - l\theta )} \right]}  \times \left( {\begin{array}{*{20}{c}}
   {{e_z}(r)}  \\
   { - i{h_z}(r)}  \\
\end{array}} \right) \\
 \end{array}
\end{equation}
Then ~$e_z(r)$~ and ~$h_z(r)$~ satisfy the Bessel's equation
\begin{equation}
\left[ {\frac{{{d^2}}}{{d{r^2}}} + \frac{1}{r}\frac{d}{{dr}} + \left( {k_ \bot ^2 - \frac{{{l^2}}}{{{r^2}}}} \right)} \right]\left( {\begin{array}{*{20}{c}}
   {{e_z}(r)}  \\
   {{h_z}(r)}  \\
\end{array}} \right) = 0
\end{equation}
In our case, only ~$TM_{0n}$~waveguide modes are excited. For these modes, ~$l=0$~, and longitudinal magnetic  field is zero. In the vacuum hole (~$0<r<R_1$~), where ~$\varepsilon  = \mu  = 1$~, the
fields must be regular at ~$r=0$~, thus
\begin{equation}
{e_z}(r) = {e_z}(0){I_0}(x)
\end{equation}
where ~$I_0(x)$~ is the modified Bessel function, and ~$x \equiv \left| {k_ \bot ^{\left( 1 \right)}} \right|r$~, and ~${\left( {k_ \bot ^{\left( 1 \right)}} \right)^2} \equiv {\left( {\frac{\omega }{c}} \right)^2} - {k^2} < 0$~. In the dielectric region(~$R_1<r<R_2$~) with outer conducting boundary, where ~$\varepsilon  = \varepsilon _2 =\varepsilon  _r$~, and ~$\mu=\mu_2$~, one has ~${e_z}({R_2}) = {e_\theta }({R_2}) = {h_r}({R_2}) = 0$~, these are equivalent to simply ~${e_z}({R_2}) = d{h_z}({R_2})/dr = 0$~. Thus
\begin{equation}
{e_z}(r) = {e_z}(0){I_0}\left( {{x_1}} \right)\frac{{{E_0}(y)}}{{{E_0}({y_1})}}
\end{equation}
where~$x_1 \equiv \left| {k_ \bot ^{\left( 1 \right)}} \right|R_1$~,~${E_0}(y) \equiv {J_0}(y){N_0}({y_2}) - {N_0}(y){J_0}({y_2})$~ with ~$y \equiv \left| {k_ \bot ^{\left( 2 \right)}} \right|r$~,~$y_1 \equiv \left| {k_ \bot ^{\left( 2 \right)}} \right|R_1$~,~$y_2 \equiv \left| {k_ \bot ^{\left( 2 \right)}} \right|R_2$~and ~${\left( {k_ \bot ^{\left( 2 \right)}} \right)^2} \equiv \varepsilon \mu {\left( {\omega /c} \right)^2} - {k^2} > 0$~, the~$J_0$~and~$N_0$~are ordinary Bessel functions of the first and second kinds. The boundary condition at ~$r=R_1$~ is that~$e_z$~,~$h_z$~,~$e_\theta$~,~$h_\theta$~are continuous, then we can get the
equation
\begin{equation}
\frac{{I_{_0}^\prime({x_1})}}{{{x_1}{I_0}({x_1})}} + \varepsilon \frac{{E_{_0}^\prime({y_1})}}{{{y_1}{E_0}({y_1})}} = 0
\end{equation}
The~$n_{th}$~root of this equation is the~$k_n$~, and the ~$f_n=c\cdot k_n/2\pi$~is the frequency of the~$n_{th}$~mode radiation excited in the structure.

Then, we go to the power solution of these excited modes when a charge bunch traverses the structure, and the theory in Reference\cite{lab20} is used.
For~$N=2$~concentric dielectric layers in the uniform cylindrical waveguide, the orthonormality relation between any two modes can be written as:
\begin{equation}
\sum\limits_{i = 1}^{N = 2} {\int_{{R_{i - 1}}}^{{R_i}} {dr \cdot r} } \left[ {{\varepsilon _i}{e_{z,m}}\left( r \right){e_{z,n}}\left( r \right) + {\mu _i}{h_{z,m}}\left( r \right){h_{z,n}}\left( r \right)} \right]={C_n}{\delta _{mn}}
\end{equation}
where~$C_n$~is the normalization constant to be used when a moving charge bunch is the source of the fields. The ~$\overline {{P_{0n}}}$~ is the power radiated into the~$TM_{0n}$~mode:
 \begin{equation}
\overline {{P_{0n}}}  =  - cq_0^2\beta \frac{{e_{z,n}^2\left( 0 \right)}}{{{C_n}}}\Theta ( - s) \cdot g({\sigma _z})
\end{equation}
where ~$q_0$~is the charge,~$c\beta$~is the velocity of the electron,~$\Theta ( - s)$~means the radiation is excited behind the electron,~$g({\sigma _z})$~is the form factor. For gaussian shape, ~$g({\sigma _z}) = \exp ( - 4{\pi ^2}\sigma _z^2/\lambda _n^2)$~, where~$\sigma_z$~is the rms length of the bunch,~${\lambda _n} = 2\pi /{k_n}$~is the wavelength of~$TM_{0n}$~. For uniform shape, ~$g({\sigma _z}) = {\sin ^2}({k_n} \cdot \sqrt 3 {\sigma _z})/{({k_n} \cdot \sqrt 3 {\sigma _z})^2}$~. Fig. 7 shows the power of wakefield radiation as a function of frequency for electron bunches with different rms lengthes, where the charge is 200 pC, the beam energy is 5.55 MeV, the inner and outer radii of the structure are~$R_1=1.2~mm$~,~$R_2=6.5~mm$~respectively. The material of the dielectric is fused silica~$\varepsilon_r=3.8$~.
 \begin{center}
    \includegraphics[width=8.0cm]{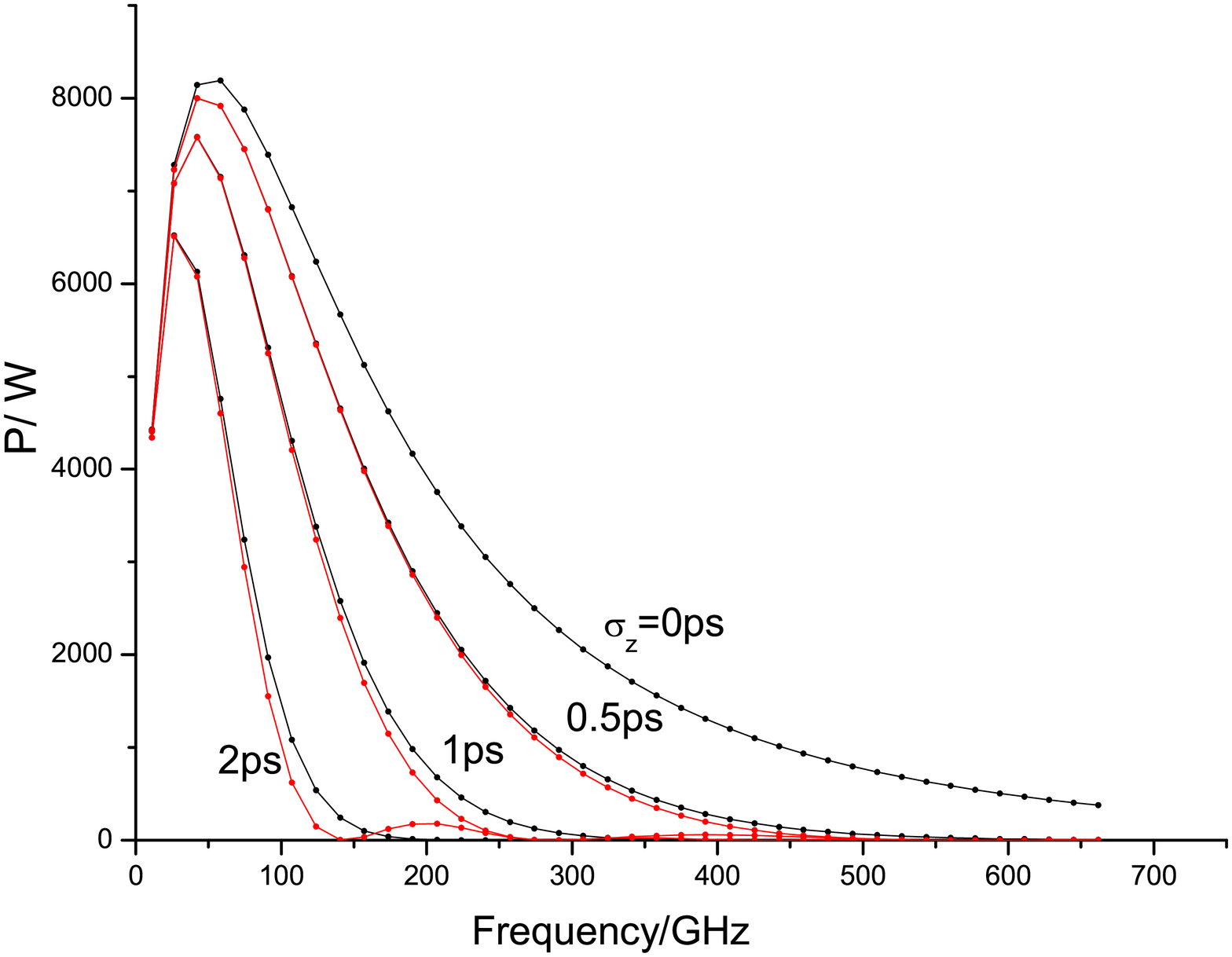}
    \figcaption{\label{fig7} The power of wakefield radiation as a function of frquency, for bunches with different rms length. The black line is for gaussian shape, and the red line is for uniform shape}
    \end{center}

The sketch of the measurement setup is shown in Fig. 8. The dielectric length should be several centimeters to make sure that the Cerenkov wakefield radiation dominates the transition radiation which is emitted as the bunch enters or leaves the structure\cite{lab21}.
\begin{center}
    \includegraphics[width=7.0cm]{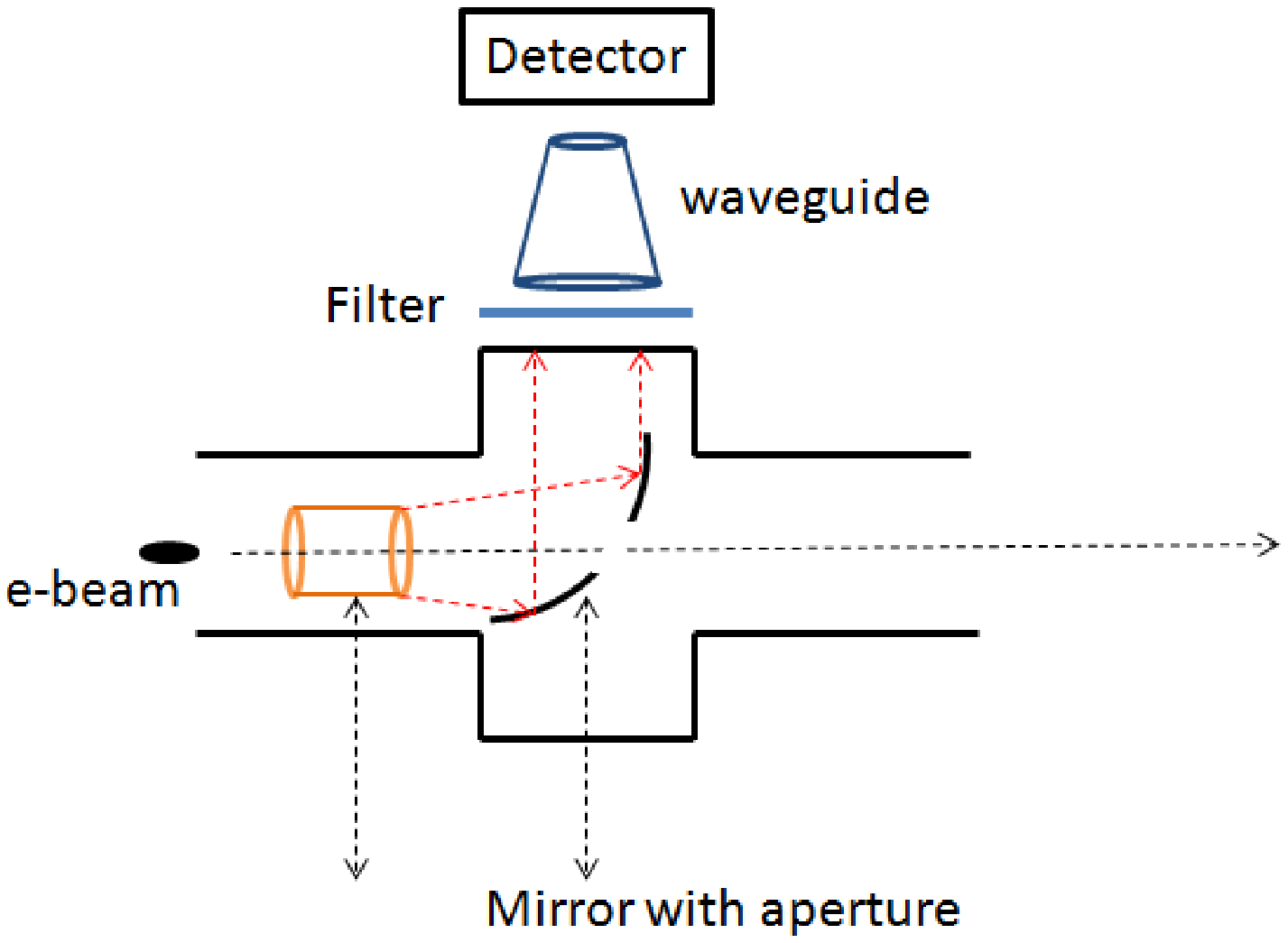}
    \figcaption{\label{fig8}Sketch of the measurement setup}
    \end{center}
The radiation emits from the structure, and then is reflected parallel by a parabolic mirror. The aperture in the mirror is used to ensure the passage of electron beam, and the transition radiation generated by the passage of electron beam through the aperture is weak and easy to calibrate. The filters filtrate out the other radiation except for the radiation at specified frequency. The detector can be a Schottky barrier diode, a golay cell, or a bolometer. The precision of the diode is relatively low compared with the golay cell and bolometer, and it works in certain bandwidth. The bolometer is expensive and needs a cryogenically cooled environment. The golay cell is a good choice with high precision and portability. Three or four filters will be used to reconstruct the spectrum of the radiation, and the bunch length can be measured. After calibration, just one filter is needed, and single shot measurement can be achieved. Besides, the charge of the bunch can also be concluded.

To ensure the passage of electron bunch, the dielectric structure should be installed around the focal point of the solenoid. The transverse beam size evolution along the longitudinal position is shown in Fig. 9.
 \begin{center}
    \includegraphics[width=7.0cm]{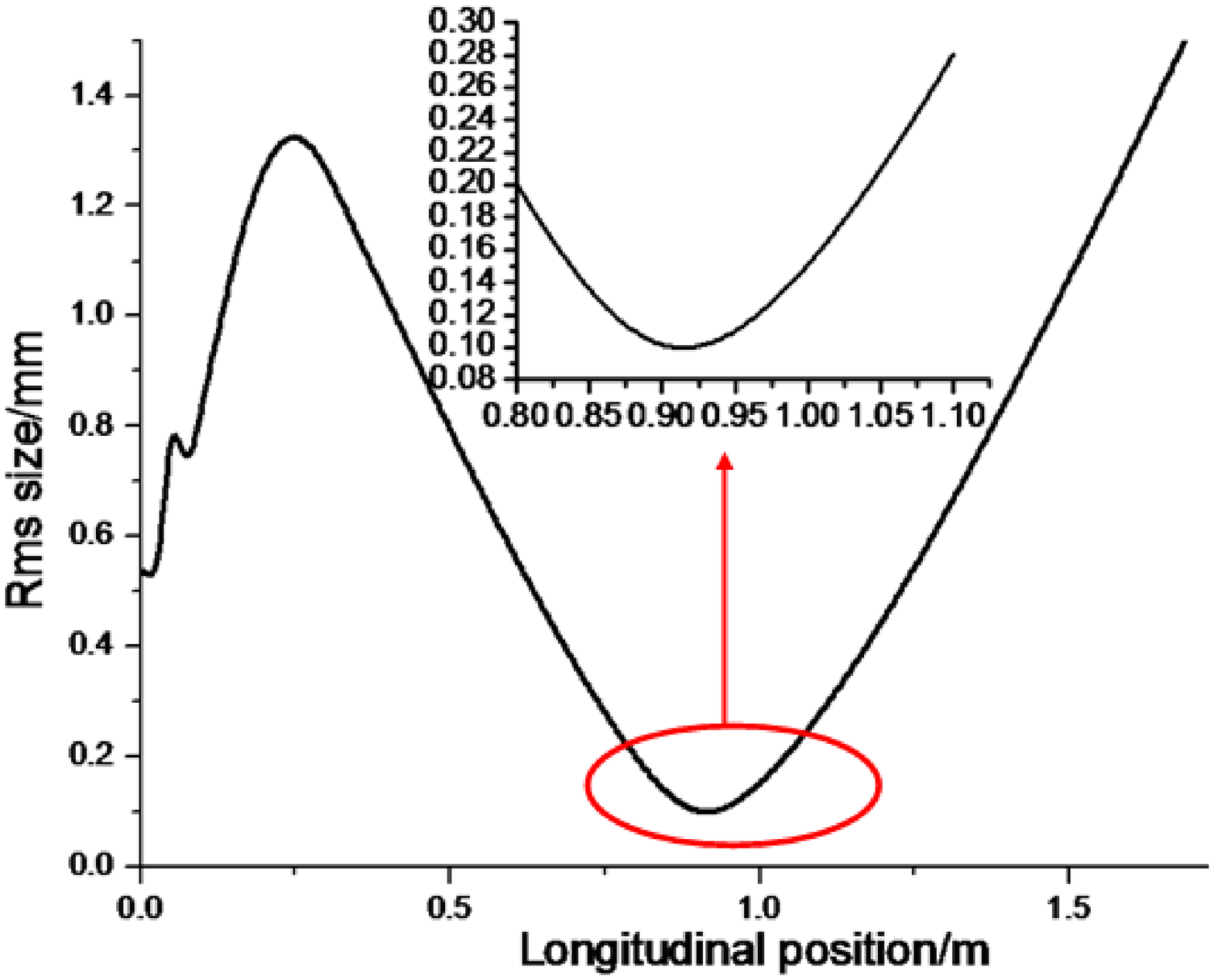}
    \figcaption{\label{fig9}Evolution of the transverse beam size along the longitudinal position}
    \end{center}
Considering the inner radius of the dielectric structure is 1.2 mm, the electron beam can pass the measurement setup in a long range without electron loss.
\section{Summary and discussion}
In this paper, we analyse the impact factors of the electron bunch length, and draw the conclusion that to generate sub-picosecond electron bunch directly in the photocathode rf gun, we need to improve the acceleration gradient of gun as high as possible, restrain the growth of transverse emittance through laser shaping technique, and carefully tune the energy (bunch charge) and diameter of the laser spot according to the requirements of applications. In order to measure the electron bunch length, the coherent Cherenkov radiation wakefield is also analysed, which is excited by the relativistic electron bunch traveling through a hollow cylindrical dielectric element. Based on the analysis, a nondestructive technique for the measurement of bunch length is reported. The advantage of this technique is routine to calibrate, and can be single shot measurement.

To replace the dielectric element used in the measurement setup with a redesigned one, narrow band THz radiation (around 0.3 THz) with high peak power (hundreds KW) can be excited, and this can be applied to table-top THz source.

\end{multicols}

\vspace{-1mm}
\centerline{\rule{80mm}{0.1pt}}
\vspace{2mm}

\begin{multicols}{2}

\end{multicols}

\clearpage

\end{document}